\documentclass[11pt]{article}
\usepackage{geometry}                
\usepackage{graphicx}
\usepackage{amssymb}
\usepackage{amsmath}
\usepackage{fixmath}
\usepackage{MnSymbol}
\usepackage{amsfonts}
\usepackage{array}
\usepackage{multirow}
\DeclareMathAlphabet{\mathpzc}{OT1}{pzc}{m}{it}

\title{Conservation Laws and Bounds on the Efficiency of Wind-Wave Growth}
\author{Clifford Chafin\\\ \small{Department of Physics, North Carolina State University, Raleigh, NC 27695} \thanks{cechafin@ncsu.edu}}

\begin{document}
\maketitle
\begin{abstract}
We examine two means by which wind can impart energy to waves: sheltering and deposition of material upwards from windward surface shear.  The shear driven deposition is shown to be the more efficient process.  Lengthening of waves to match the wind speed is shown to be very inefficient and consume a large fraction of the energy imparted by the wind.  The surface shear provides a low energy sink that absorbs most of the momentum from the wind.  These produce bounds on the efficiency of wave growth.  The results here are computed in a model independent and perturbation free fashion by a careful consideration of conservation laws.  
By combining these effects we can place bounds on the rates waves can grow in a given fetch and the relative amount of shear flow versus the, relatively small, Stokes drift that must arise.  
\end{abstract}

The generation of waves by wind is still not completely understood.  It is known that wind blowing over a flat surface first generates instabilities that leads to progressive capillary waves that move at oblique angles until the component of their motion parallel to the wind matches the wind speed.  When the waves become large enough, instabilities create longer waves and the pressure difference on the windward and leeward side lead to ``wave sheltering'' that imparts energy and linear momentum to the waves \cite{Ca07}.  Since these forces appear by surface pressure changes and this, to lowest order, gives pressure fields that are solutions to Laplace's equations, the forces and velocity fields induced are irrotational in a constant density fluid.  Irrotational waves carry momentum through Stokes drift so the energy imparted and momentum imparted must be matched.  This is generally not possible for a fixed shape waveform except for very particular pressure fields.  

Some concerns here are what happens with angular momentum.  It has been shown recently that torques on water surfaces are typically carried out to the ends of long packets and that infinite waves on flat surfaces have boundary condition problems with respect to conservation laws because of this \cite{Chafin-rogue}.  Another confounding point is that surface waves typically have deeper ``Eulerian'' flows \cite{Smith} that seem to cancel the Stokes drift so that no net momentum flow (mass flux) from the waves exists.  We will show that the efficiency of the process is low enough so that the Stokes drift will generally be swamped by the large surface shear created by the wind.  
Our goal is not to provide a unifying theory of wind-wave interaction but to carefully utilize conservation laws and surface forces to give some universal bounds on the efficiency and mechanisms of wave growth.  

This paper is organized as follows.  A discussion and criticism of the perturbative theory of wave interaction and equilibration is given.  Following this is a comparison of wave sheltering versus shear driven ``crest deposition'' means of wave growth.  Conservation laws are summarized and angular momentum is used to give bounds on the losses to surface shear and heating as the wavelength of the mean surface waves grow.  Efficiency bounds are given considering the energy losses to wave crest destruction and viscous surface losses. 

\section{Wave Interactions}
Waves rarely occur in a nearly ``monochromatic'' distribution.  There is often a broad distribution of waves especially in the period of rapid growth.  Nonlinearities are typical during this and sharply peaked waves and breaking is common.  The interaction of these waves is believed to be important in creating an ``equilibrium'' distribution over a given fetch and duration of wind.  

The Navier-Stokes equations
\begin{align*}
\frac{\partial}{\partial t}v+v\cdot\nabla v=-\frac{1}{\rho}P+gz+\nu\nabla^{2}v
\end{align*}
can be linearized for fluid at rest on the bottom with irrotational surface waves described by the basis of pairs
\begin{align}
\phi_{k}(x,z)&=e^{kz}\cos(kx)\\
\eta_{k}(x,t)&=\frac{k}{\omega}\phi_{k}(x,0)\sin(\omega t)\\
\phi_{k}^{\dagger}(x,z)&=e^{kz}\sin(kx)\\
\eta_{k}^{\dagger}(x,t)&=-\frac{k}{\omega}\phi_{k}(x,0)\cos(\omega t)
\end{align}
where $\phi$ is the velocity potential and $\eta$ is the surface deformation.  These linear wave solutions are called Airy waves \cite{Ai45, Stokes} but superposition here is not as free as for linear superpositions in optics and some other wave theories.  This is due to the nonlinear term in the Navier-Stokes equations.  

The wavevector $k$ can be described in terms of the wavelength so such eigenstates can be described by the pair $a,\lambda$ and the direction of propagation.
Given waves with amplitudes and wavelengths $a,\lambda$ and $A,\Lambda$, when these are close in value we see that the nonlinear term is small in the sense that we can decompose the N-S equation into two independent equations for each component.  However, when $\lambda\ll A$ this is not true and the nonlinear interaction term is comparable or larger than the linearized part corresponding to the wave $a,\lambda$.  Furthermore, given a distribution of small wave components, there can be locations and times where the steepness of the wave is so large that breaking occurs.  Thus the ``smallness'' condition in hydrodynamics is not as trivial as in the case of relativistic free field theories.  We can always represent the surface deformation of nonbreaking waves on the basis of $\eta_{k}$ and $\eta_{k}^{\dagger}$.  However, the pressure distribution of has a nonlinear correction and is a function of the changing surface so that a description in terms of the above basis is not sufficient.  The interaction of much smaller waves on larger ones $a\ll A$ means they move across the surface of the larger ones so that a superposition of the pressure fields is very inaccurate.  Given that the superposition of such waves is not accurate over the time scale of a single period $\Omega$ of the larger wave, it is hard to see how a perturbative theory of interactions that build up gradually over time can be useful.

%

One way people have attempted to study wave interactions is through perturbation theory drawn from Feynman diagrams and Quantum Field Theory \cite{Ha}.  The problem with this approach is that, while it is a theory of scattering, it works by assuming a 1-1 correspondence of the interacting and noninteracting states at infinity.  The ``adiabatic'' turn off of interactions allows the dressing and interaction of the particles to occur.  These interactions are always linear.  The complications come from many body contributions that are difficult to include in the initial data.  Surface waves come from a fundamentally nonlinear equation of motion and the interactions can lead to breaking that introduces vorticity at the surface and flows that are not represented in our basis set.  A recent discussion of how the Stokes perturbation expansion is inadequate to describe the long range pressure fields of wave packets is given by Chafin \cite{Chafin-rogue}.


Observations indicate that large waves tend to be very destructive to small amplitude waves.  The mechanism for this is not entirely clear but an advancing set of waves do not tend to allow a smaller set of waves to traverse against them for long.  The energy and momentum of these must be lost to heat and surface flow.  For now we take this as a given effect and investigate the role of external forces on waves.

\section{Crest Deposition and Wave Sheltering}
Consider the following thought experiment.  A uniform propagating wave $A,\Lambda$ is moving the the right.  If we now gradually add mass to the top of the crests and remove it from the troughs in a sinusoidal fashion, then the waves increase in kinetic energy but there is no bias in the direction of this change so we increase our rightwards wave to $A+\delta A,\Lambda$ and introduce a backwards wave $\delta A,\Lambda$.  These give a net zero linear and angular momentum contribution.  The energy is divided up between these waves.  
Since we presume the small backwards waves are destroyed by surface instabilities, this process is, at best, 50\% efficient.  This process is adding energy to the potential part of the wave, not the kinetic portion directly.  The fluid could be added at the local velocity of the wave to improve efficiency but there is no way to enhance the patter of deep flow for a purely propagating wave directly this way.  The increased potential energy converts to velocity fields of a standing wave component.  

In a realistic case of wind generated waves we expect some pressure field that is enhanced on the windward side and weakened on the leeward one.  This provides forces in the direction the water in the wave is locally moving so should impart energy to it.  
One might wonder if these two processes could work together to be more efficient or if one dominates.  We will investigate this next.  

\begin{figure}[!h]
   \centering
   \includegraphics[width=4in,trim=0mm 170mm 0mm 170mm,clip]{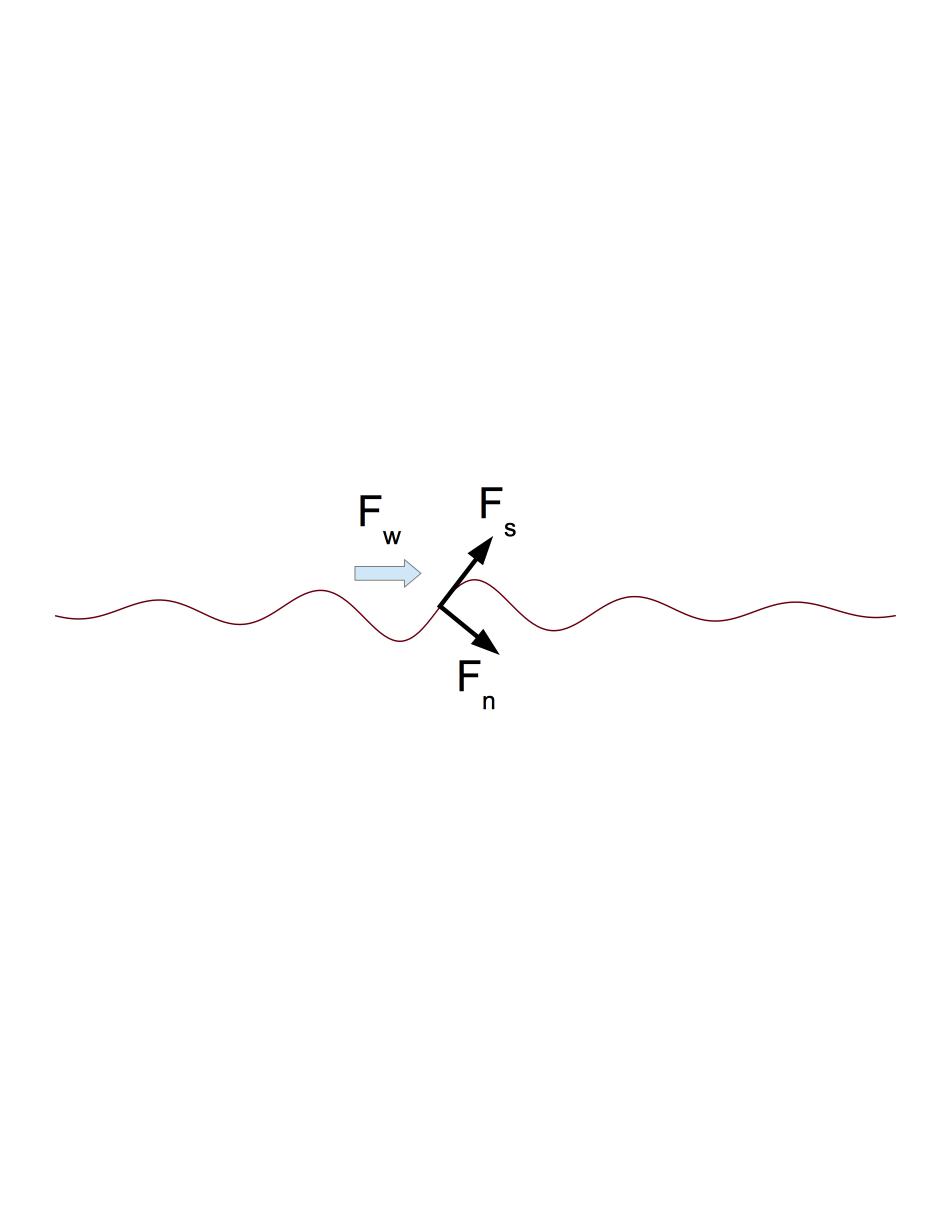} 
   \caption{Wind force on surface.}
   \label{tangents}
\end{figure}

\begin{figure}[!h]
   \centering
   \includegraphics[width=4in,trim=0mm 170mm 0mm 220mm,clip]{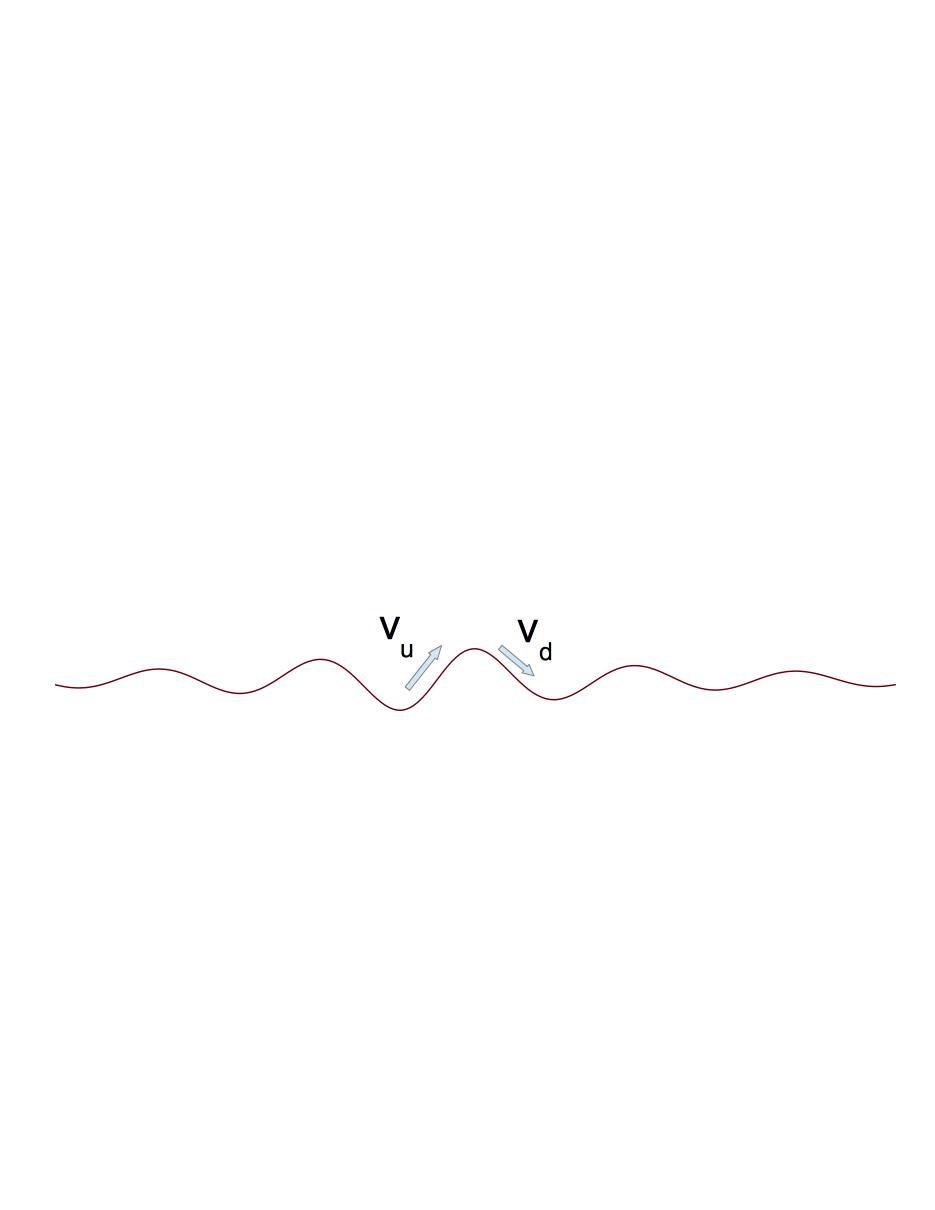} 
   \caption{Wind speed on upwards and downwards slope of wave.}
   \label{winds}
\end{figure}

We can quantify both processes by specifying the net shear drag on the surface from the wind.  This is different than the usual process of specifying the specifics of the pressure field. The turbulent air surface flow is an easier to measure quantity and can be used to predict this surface impulse on the water.  This force is horizontal and can be approximated by the diagram in fig.\ \ref{tangents}.  
As the wind moves over the wave that is moving less than the wind speed, the wind flows faster up the windward than the leeward side.  For turbulent flow, the leeward wind may be a turbulent eddy that moves backwards to the wave as in fig.\ \ref{winds}.  
We expect the no slip conditions to hold but the wave surface may also be dragged with the wind so that it is not at rest with the water beneath it.  If we define the wind speeds $v_{u}$ and $v_{d}$ to be just above the surface so that there is no turbulence at this level then we can ask how much fluid it drags up with it.  If $v_{d}$ is very small, as when $v_{\text{wave}}\ll v_{\text{wind}}$, then we can consider these to be periodic pulses of shear flow that are then equilibrated to zero on the leeward side.  The boundary layer of liquid dragged is then laminar up to a separatrix which forms on the leeward side of the wave.  The thickness of laminar boundary layer is know from
\begin{align}
\delta=&\frac{5 l}{\sqrt{Re}}\\
Re=&\frac{v_{s} l}{\nu}
\end{align}
The length of flow drag when $v_{s}\ll v_{wave}$ is $l\approx v_{s}\frac{T}{2}=v_{s}\frac{\pi}{\omega}$ where $\omega$ is the wave frequency, $T$ is the wave period and $v_{s}\approx v_{u}-v_{d}$ is not the averaged surface shear but the transient one set up on windward side of the wave.  The Reynolds number, $Re$, here is that of the transient surface flow in the liquid.  
The rate mass is transported up the wave is $\delta\rho v_{s}$.  The mass dragged up the wave will take $n=\frac{\lambda/2}{l}$ periods to reach the top.  This tells us that the rate of wave growth is 
\begin{align}
\dot{a}_{cd}=\frac{2\delta l}{T\lambda}\approx \frac{\delta v_{s}}{\lambda}
\end{align}
Given that wave periods are on the order of a few to 10's of seconds and surface flow velocity are less than a m/s, this gives skin depths of mm's and growth rates of $\sim10^{-5}$m/s from this effect.  Consulting wave fetch tables we see that this is of the right order of magnitude for realistic wave growth.  

When the wave speed approaches the wind speed, we expect $v_{u}\approx v_{d}$ so that the drag force vanishes.  The drag force on the surface still exists but now transfers all of its energy to the surface shear flow.  

Now let us estimate the growth from wave sheltering.  At the surface of the fluid the air flow is laminar in a thin layer.  This allows the pressure field at the surface of the wave can be estimated by Bernoulli's principle $\rho_{air} v_{s}^{2}/2+P=$~const.  For wind blowing in the same direction as the waves and slightly faster than them this is a positive increase in energy.\footnote{Interestingly, this model should reduce the size of waves when wind blows against them yet this is exactly the condition where we tend to find massive ``rogue'' waves.  The crest deposition mechanism is enhanced in this situation.  }  Thus we expect pressure differences between the windward and leeward sides of the wave as less than a Pa/m$^{2}$.  The force on shallow water waves is almost perfectly downwards and the power imparted to each wave crest is 
\begin{align}
\dot{{E}}\approx a\omega \Delta P \frac{\lambda}{2}
\end{align}
The energy density of a wave is $\mathcal{E}=\frac{1}{2}\rho_{w} g a^{2}$ so the power imparted per crest is
\begin{align}
\dot{{E}}\approx \rho_{w} g a \dot{a}\frac{\lambda}{2}
\end{align}
This yields a growth rate
\begin{align}
\dot{a}_{sh}\approx \frac{\omega \Delta P}{\rho_{w} g} 
\end{align}
We have not included an efficiency factor to this process because of the following.  
The conserved quantities in waves are expressed\footnote{The mass density $\mathit{m}$ is explained in \cite{Chafin-rogue} as due to pressure driven elevation effects.} in the table \ref{tab}. 
\begin{table}
\begin{center}
\vspace{0.5cm}
\begin{tabular}{|l|l|l|l|}  \hline $\mathcal{E} $& $\mathit{m}$&$\mathpzc{p}$ & $\mathcal{L}$ \\  \hline \multirow{2}{*}{$\frac{1}{2}\rho g a^{2}$}& \multirow{2}{*}{$\rho a^{2}  k$}  & \multirow{2}{*}{$\frac{1}{2}\rho a^{2} \omega $} &\multirow{2}{*}{$-\frac{1}{4}\rho g \frac{ a^{2}}{\omega}$}  \\ &  &  &\\ \hline \end{tabular} 
\vspace{0.5cm}
\end{center}
\caption{Conserved depth-integrated and time averaged quantities.}\label{tab}
\end{table}
By comparing the momenta to energy density of the waves we see
\begin{align}\label{pE}
\frac{\mathpzc{p}}{\mathcal{E}}&=\frac{k}{\omega}
\end{align}
However, comparing with the ratio of the force to the power
\begin{align}\label{fPow}
\frac{f}{\dot{\mathcal{E}}}&=\frac{k}{\omega}
\end{align}
the momentum and energy imparted are equivalent so that we expect the efficiency to be large.\footnote{We have not explicitly included angular momentum here but by \cite{Chafin-rogue} only the $\tau_{z}=z F_{x}$ contributions will be interesting to us in altering our local wavetrain and not expelled at the end of a packet.}  

We can compare the effects on elevation growth of the wave from sheltering and crest deposition by using the Bernoulli estimate for the pressure
\begin{align}
\dot{a}_{sh}\approx \frac{1}{2}\frac{\rho_{a}}{\rho_{w}}\frac{\omega}{g}v_{s}^{2}=\frac{1}{2}\frac{\rho_{a}}{\rho_{w}}v_{ph}^{-1}v_{s}^{2}
\end{align}
and the skin depth estimate for $\delta$
\begin{align}
\dot{a}_{cd}\approx 5\pi\frac{v_{s}^{2}}{\lambda\omega}Re^{-1/2}=\frac{5}{2}v_{ph}^{-1}v_{s}^{2}Re^{-1/2}
\end{align}

We can compare the relative sizes of these contributions by
\begin{align}
\frac{\dot{a}_{sh}}{\dot{a}_{cd}}=\frac{\frac{\rho_{a}}{\rho_{w}}}{5Re^{-1/2}}
\doteq\frac{1.2\times10^{-6}}{5\times 10^{-3}}\sim10^{-3}
\end{align}
so that crest deposition generally plays the dominant role.

\section{Wave Lengthening}
We now have some measure of the growth rate under sheltering and crest deposition in terms of the surface shear force of the wind.  The net momentum of the waves and shear flow is simply $\int F_{w} dt$.  This of course disguises that the surface force is not just a function of the winds but the details of wave shape and motion that generate traction on the upper level winds.  The energy imparted to shear flow ends up primarily as heat as the current diffuses into deeper waters.  The wave energy is a more persistent quantity as low viscosity ensure long waves may go thousands of miles with small attenuation.  

The model we introduced above increased amplitude but never altered the wavelengths of the ambient waves.  Wave interaction models are designed to express how waves can lengthen and pump energy down to smaller waves where instabilities can destroy them \cite{Ha}.  This provides a way to increase the speed and length of the observed waves.  Since we have expresses some doubt in the validity of these results let us seek another approach that involves independent assumptions.  

As the wind blows over a sufficiently long isolated fetch of sea for a time that is much less than the wave crossing time of the region we can assume that all wave growth, current generation and wave destruction are confined to this region.  Very long fast waves may escape it and end-of-packet regions may absorb large net quantities of angular momentum due to long range pressure variation of the driving parcel of air above it \cite{Chafin-rogue}.  Beyond that a quasilocal description of the waves based on conserved quantities seems valid.  Let us presume we do not know the details of how the waves lengthen and approach the wind speed and that there is a dominant wave amplitude and wavelength $a, \lambda$ that describe the swells and their conserved quantities.  Since the momentum and angular momentum of the waves are sensitive to the frequency a broad distribution weakens the validity of these assumptions.  With that in mind we continue to develop this ``monochromatic model'' of wave development.  

The region of sea that the wind blows over has some long fetch $L\gg \lambda$.  There is some spreading and energy loss out the ends of this but if $L$ is large enough and the blow up time for the waves is short compared to the crossing time of the waves we can assume that the local angular momentum of the waves is conserved.  The damping of an irrotational flow due to viscosity must be a singular contribution at the free surface.  Since there is no lever arm to generate a torque here, this contributes to the linear momentum of the shear flow but cannot transfer angular momentum to it.   
This implies there is a process of crest and trough ``destruction'' as the number of waves in the given space are reduced.  From
\begin{align}
\frac{1}{4}\rho \frac{a^{2}}{\omega}=\text{constant}
\end{align}
we have
\begin{align}
\dot{a}=\frac{1}{2}a \frac{\dot{\omega}}{\omega}
\end{align}
or that the energy loss from heating is 
\begin{align}
\mathcal{P}_{d}=\mathcal{E}\frac{\dot{\omega}}{\omega}
\end{align}
where $\mathcal{E}=\frac{1}{2}\rho g a^{2}$ is the energy density of the waves.  

\section{Efficiency}

Assume that the wave amplitude and wavelengths evolve in time as $a(t), \lambda(t)$ in response to the surface forces $F(t)$ (which is an implicit but unspecified function of the waves themselves).  If there is no surface shear to begin with, the momentum of the wave and shear give the full impulse of the wind.  Energy is not conserved for the shear flow.  For the waves, energy is conserved until some destructive event occurs.  To lengthen waves they must gain angular momentum or lose amplitude.  The momentum of the waves decreases as waves length for fixed amplitude.  This momentum loss goes directly into shear flow.  Angular momentum can be made up for with end-of-packet elevation changes.  To give a maximally efficient process of wave growth consider the loss of linear momentum to indicate the energy loss of a wave.  

The power (per area) imparted to the sea from the wind is 
\begin{align}
\mathcal{P}_{w}\approx\frac{\omega \Delta P}{\rho_{w} g} + F_{w}v_{s}+(\rho\delta)\cdot g\cdot \left(v_{s}\frac{a}{\lambda}\right) 
\end{align}
The losses to shear motion are
\begin{align}
\mathcal{P}_{s}\approx F_{w}v_{s}\approx&\mu \frac{v_{s}}{\delta}v_{s}\\
=&\frac{1}{5}\sqrt{\frac{\mu\rho}{\pi}}\omega^{1/2}v_{s}^{2}\doteq0.11\omega(t)^{1/2}v_{s}(t)^{2}
\end{align}
To compute the efficiency of wave growth we compute the net wave energy relative to the net power imparted   
\begin{align}
\epsilon_{\text{eff}}=&\frac{\frac{1}{2}\rho g a_{f}^{2}}{\frac{1}{0.5}\frac{1}{2}\rho g a_{f}^{2}+\int dt~(\mathcal{P}_{s}+\mathcal{P}_{d})}\\
\approx&\frac{\frac{1}{2}\rho g a_{f}^{2}}{\rho g a_{f}^{2}-\frac{1}{2}\rho g \int dt~a(t)^{2}\frac{\dot{\omega}(t)}{\omega(t)}
+0.11\int dt~\omega(t)^{1/2}v_{s}(t)^{2}}\\
<&\frac{\frac{1}{2}\rho g a_{f}^{2}}{\rho g a_{f}^{2}-\frac{1}{2}\rho g \int dt~a(t)^{2}\frac{\dot{\omega}(t)}{\omega(t)}}\\
=&\frac{1}{2}\frac{ a_{f}^{2}}{ 2a_{f}^{2}- \int dt~a(t)^{2}\frac{\dot{\omega}(t)}{\omega(t)}}
\end{align}
As noted, the growth of the wave by crest deposition is at best 50\% efficient from shear forces justifying the inclusion of the $0.5$ efficiency factor in the denominator.  

For waves that grow roughly linearly in amplitude and frequency this gives an upper bound on the efficiency of 
\begin{align}
\epsilon_{\text{eff}}\le \frac{1}{2}\left(\frac{1}{2+\frac{1}{3}}\right) = \frac{3}{14}
\end{align}
More detailed wave growth measurements can then yield more accurate efficiencies.  In practice, measuring the efficiency requires measuring the detailed traction force of the wind on the waves or using a calorimetric measurement of heat and the small remaining KE of shear flow.  Such a value is not really that valuable. However, the momentum imparted to the wave versus the shear flow is interesting and relevant given the long lasting issues with resolving Stokes drift in waves.  The efficiency of wave growth is related since the energy density and momentum of the waves are proportional by a factor of $\omega$.  We can see that the efficiency of momentum transfer to the wave motion versus the shear flow is instantaneously related to the angle of the wave surface $\theta\approx\arctan(a/\lambda)$.  This shows that, even without wave destruction, the vast majority of momentum is imparted to the shear flow.  This is important since, if most of the energy is to be imparted to the waves, there must be a large momentum sink that consumes relatively little of the energy.  

\section{Conclusions} 
We have shown that the dominant process that drives wave growth is probably the elevation of mass through wind shear from lower regions to the crests rather than sheltering.  The role by which waves lengthen their periods is not clear but we have assumed no specific wave growth model.  The derivations are based on a careful consideration of the conservation laws so the bounds established are quite general.  It has been unclear, and generally not considered, how waves which carry so little momentum yet so much energy can gain these so disproportionately from the wind which applies a net transverse traction force on the surface.  Proper inclusion of the shear drag is essential here and bounds on efficiency of wave growth follow.  

Previous calculations involving waves typically follow from the precomputer days of analytic tour de force where symmetry arguments, idealized geometries and the aesthetic appeal of the results was a strong case for their validity.  This has some celebrated successes but nature is not always so kind.  There are physical processes which are just ugly and subtle by nature.  Some involve such small scale effects that computers may not be able to model them without great insight on the part of the practitioners.  The arguments here suggest that wave growth is such a process and not just because of turbulence induced in the driving air stream.  The boundary layer effects as transient biased elevators of fluid to build waves will generally be confined to layers millimeters thick with waves whose irrotational flow depths will be 10's of meters.  The feedback between wind and wave motion will be irregular and essential.  Finding optimal mixtures of theoretical analysis and computation to arrive at insight, understanding and quantitative work will likely remain a formidable challenge.  It is this author's belief that this sort of discussion can filter out enough of the complexity while retaining the salient features required for less approximate results and point the way to more potent computer simulations.

\end{document}